\documentstyle[svcon,psfig]{report} 
\newcommand{\lesssim}{\raise.3ex\hbox{$<$\kern-.75em\lower1ex\hbox{$\sim$}}}
\newcommand{\gtrsim}{\raise.3ex\hbox{$>$\kern-.75em\lower1ex\hbox{$\sim$}}}
\begin{document} 
\pagenumbering{arabic}
\chapter{Scale competition in nonlinear Schr{\"o}dinger models}
\chapterauthors{Yu.B.~Gaididei \\
P.L.~Christiansen \\
S.F.~Mingaleev}
\begin{abstract} 
Three types of nonlinear Schr{\"o}dinger models with multiple length
scales are considered. It is shown that the length-scale competition
universally results into arising of new localized stationary states.
Multistability phenomena with a controlled switching between stable
states become possible. 
\end{abstract}  

\section{Introduction} 

The basic dynamics of deep water and plasma waves, light pulses 
in nonlinear optics and charge and energy transport in condensed 
matter and biophysics \cite{cit1,cit2,cit3,cit4} is described by the 
fundamental nonlinear Schr{\"o}dinger (NLS) equation
\begin{equation}
i\frac{\partial}{\partial\,t} \psi+L^2\partial^{2}_x\psi+ 
V|\psi|^{2}\psi+f(x)\psi=0 \; , 
\end{equation}
where $\psi(x,t)$ is the complex amplitude of quasi-monochromatic 
wave trains or the wave function of the carriers. The second term 
represents the dispersion and $L$ is the dispersion length (e.g. 
in the theory of charge (energy) transfer $L^2=\hbar^2/2m$ with 
$m$ being an effective mass). The nonlinear term, $V|\psi|^{2}\psi$, 
describes a self-interaction of the quasiparticle caused  either by 
its interaction with low-frequency excitations (phonons, plasmons, 
etc.) \cite{cit9} or by the intensity  dependent refractive index of 
the  material (Kerr effect) \cite{cit10}. The function $f(x)$ is a 
parametric perturbation which can be a localized impurity potential, 
a disorder potential, a periodic refractive index, an external 
electric field, etc. It is well known that as a result of competition 
between dispersion and nonlinearity  nonlinear waves with properties 
of particles, solitons, arise. One can also say that this competition 
leads to the appearance of the new length-scale: the width of the 
soliton $\zeta=L/\sqrt{V}$. The presence of the parametric 
perturbation $f(x)$ introduces additional interplays between nonlinearity, 
dispersion and perturbations. In the recent paper by Bishop {\em et al.} 
\cite{cit5} the concept of competing length-scales and time-scales 
was emphasized. In particular, Scharf and Bishop \cite{Scharf1,Scharf2} 
have discussed the effects of a periodic potential 
($f(x)=\epsilon \cos(2\pi x/\zeta_p)$) 
on the soliton of the NLS equation, and shown on the basis of an 
averaged NLS equation that for $\zeta_p/\zeta \lesssim 1$ or 
$\gtrsim 1$ the periodic potential leads to a simple renormalization 
of the solitons and creates a 'dressing' of the soliton. But when 
$\zeta_p \sim \zeta$ there is a  crucial length-scale competition 
which leads to the destruction of the soliton. Another interesting 
example of the length-scale competition was provided by 
Ref.\ \cite{cit6} where the authors showed that the propagation of 
intense soliton-like pulses in systems described by the 
one-dimensional NLS equation may be left practically unaffected by 
the disorder (when $f(x)$ is a Gaussian $\delta$-correlated process). 
This theoretical prediction has recently been confirmed 
experimentally using nonlinear surface waves on a superfluid helium 
film \cite{cit7}.

The goal of this paper is to extend the concept of the length-scale 
competition to the essentially non-integrable systems: to systems 
with nonlocal dispersion and to systems with unstable stationary 
states.

\section{Excitations in nonlinear Kronig-Penney models} 

Wave propagation in nonlinear photonic band-gap 
materials  and in periodic nonlinear dielectric superlattices 
\cite{phot1,phot2} consisting of alternating layers of two 
dielectrics: nonlinear and linear, is governed  by the NLS equation
\begin{eqnarray}
i\partial_t \psi(x,z,t) &+& \zeta^2\,(\partial^2_x+\partial^2_z) 
\psi(x,z,t) \nonumber \\
&+& w\,\sum_n\,\delta (x-x_n)|\psi(x,z,t)|^2\psi(x,z,t)=0 \; , 
\label{eq0}
\end{eqnarray}
where $x_n=n\,\ell$ is the coordinate of the n-th nonlinear layer 
($\ell$ is the distance between the adjacent nonlinear layers), 
and it is assumed that the width  $w$ of the nonlinear layer is 
small compared to the soliton width $\zeta$  within the layer. In 
this case the problem can be described by the nonlinear Kronig-Penney 
model given by Eq.\ (\ref{eq0}). It was shown in Ref.\ \cite{gaid} that 
the field $\psi(x,z,t)$ can be expressed in terms of the complex 
amplitudes $\,\psi_n(z,t)\equiv \psi(x_n,z,t)$ at the nonlinear 
layers. The complex amplitudes $\psi_n(z,t)$ can be found from the 
set of pseudo-differential equations
\begin{eqnarray}
\frac{\zeta^2 \hat{\kappa}}{\sinh{\ell\hat{\kappa}}} 
(\psi_{n+1}+\psi_{n-1})- \frac{2 \zeta^2 \hat{\kappa}}{ 
\tanh{\ell\hat{\kappa}}}\,\psi_n + w |\psi_n|^2 \, \psi_n =0 
\label{eq5}
\end{eqnarray}
with periodic boundary conditions $\psi_{n+N}=\psi_n$, where $N$ is 
the number of layers. In Eq.\ (\ref{eq5}) the operator $\hat{\kappa}$ 
is defined as 
$\hat{\kappa}\psi= \zeta^{-1} \sqrt{-i\,\partial_t-\zeta^2\, 
\partial_z^2}\psi$. 
Equation (\ref{eq0}) has an integral of motion -- the norm (in nonlinear 
optics this quantity is often called the power) 
$P=\int_{-\infty}^{\infty} |\psi|^2 dx dz$.

For the excitation pattern where the complex amplitudes are the same 
in all nonlinear layers, $\psi_n(z,t)=\Psi(z,t)$, we get 
\begin{eqnarray}
\sqrt{-i\partial_t-\zeta^2\,\partial^2_z}\,\tanh\left( 
\frac{\ell}{2\zeta}\sqrt{-i\partial_t - \zeta^2 \, \partial^2_z} 
\right)\Psi-\frac{w}{2\zeta}|\Psi|^2\Psi=0 \; . 
\label{eq7}
\end{eqnarray}
Equation (\ref{eq7}) clearly shows the existence and competition of 
two characteristic length-scales: the interlayer spacing $\ell$ and 
the size of the soliton in the nonlinear layer $\zeta$. When 
$\ell\ll \zeta$ one can expand the hyperbolic tanhence and 
Eq.\ (\ref{eq7}) takes the form of usual NLS equation 
\begin{eqnarray}
\left(i\partial_t+\zeta^2\,\partial^2_z\right)\Psi+ 
\frac{w}{\ell} |\Psi|^2\Psi=0 \; .
\label{eq8}
\end{eqnarray}
In the opposite limit, when $\ell\gg\zeta$ and $\tanh\left( 
\frac{\ell}{2\zeta}\sqrt{\cdots}\right)\simeq 1$, equation 
(\ref{eq7}) takes the form
\begin{eqnarray}
\sqrt{-i\partial_t-\zeta^2\,\partial^2_z}\Psi- 
\frac{w}{2\zeta} |\Psi|^2\Psi=0 \; ,
\label{eq9}
\end{eqnarray}
which reduces, for static distribution ($\partial_t \Psi=0$), to the 
nonlinear Hilbert-NLS equation recently introduced in 
Ref.\ \cite{gmcr}. It is noteworthy that in contrast to usual NLS 
solitons, the localized solutions of the nonlinear Hilbert-NLS 
equation have  algebraic tails \cite{gmcr}.

It is worth to note the close relation of the problem under 
consideration to the theory of the long internal gravity waves in 
a stratified fluid with a finite depth $h$ (see e.g. \cite{kkd}) 
which are described by the equation
\begin{eqnarray}
\partial_t u+\frac{1}{h}\partial_x u+2u\partial_x u+
T\partial^2_x u=0 \; , 
\end{eqnarray}
where $T(\cdot)$ is the singular integral operator given by
\begin{eqnarray}
(Tf)(x)=\frac{1}{2h} \, p.v. \!\!\! \int\limits_{-\infty}^{\infty} 
\coth \left( \frac{\pi(y-x)}{2h} \right) f(y) dy
\end{eqnarray}
($p.v.$ means the principal value integral). In the shallow 
water limit ($h\rightarrow 0$) the dynamics is described 
by the Korteweg-de-Vries equation, 
$\partial_t u+\frac{h}{3}\partial^3_x u+2u\partial_x u=0$,
while the Benjamin-Ono equation, 
$\partial_t u+H\partial^2_x u+2u\partial_x u=0$,
governs the water wave motion in the deep-water limit 
($h\rightarrow \infty$). Here 
$(Hf)(x)=\frac{1}{\pi} \, p.v. \!\!\! \int 
\limits_{-\infty}^{\infty} dy f(y)/(y-x)$ 
is the Hilbert transform.

\begin{figure}[h]
\begin{center}
\leavevmode
\psfig{figure=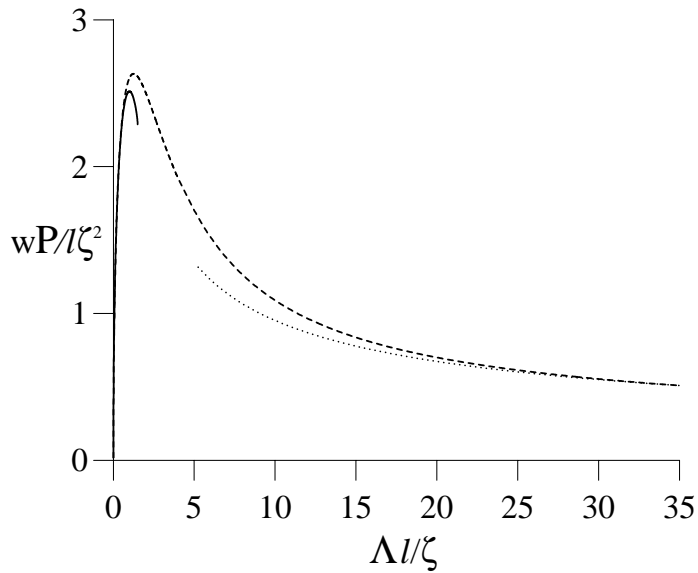,height=3.0in,angle=270}
\caption{Power $P$ of the stationary state 
$\psi_n(z,t)= e^{i\Lambda t} \phi(z)$ vs the nonlinear frequency
$\Lambda$. Numerical results from Eq.\ (\ref{eq11}) 
(dashed line), Pad{\'e} approximation (full line) and the 
asymptotic relation $P \sim \Lambda^{-1/2}$ as 
$\Lambda \to \infty$ (dotted line).} 
\label{fig1}
\end{center}
\end{figure} 

Being interested in stationary states of the system we consider 
solutions of the form $\Psi(z,t)= \phi(z) \exp(i \Lambda t)$, 
where $\Lambda$ is the nonlinear frequency and $\phi(z)$ is the 
real shape function. Since Eq.\ (\ref{eq5}) is Galilean invariant, 
standing excitations can always be Galileo boosted to any velocity 
in $z$-direction. For the shape function $\phi(z)$ we obtain a 
nonlinear eigenvalue problem in the form
\begin{eqnarray}
\sqrt{\Lambda -\zeta^2\,\partial^2_z}\,\tanh\left( 
\frac{\ell}{2\zeta}\sqrt{\Lambda - \zeta^2\,\partial^2_z} 
\right)\phi-\frac{w}{2\zeta}\phi^3=0 \; . 
\label{eq11}
\end{eqnarray}
Simple scaling arguments show that in the low-frequency limit 
($\Lambda \ell^2/ \zeta^2 \rightarrow 0$) the norm behaves in 
the same way as in the case of usual NLS equation (\ref{eq8}): 
$P\sim \sqrt{\Lambda}$. When $\Lambda \ell^2/ \zeta^2\rightarrow 
\infty$ the norm $P$ is a monotonically decreasing function: 
$P\sim 1/ \sqrt{\Lambda}$. From the analysis of Ref.\ \cite{gaid} 
follows that the norm $P(\Lambda)$ is a nonmonotonic function with 
a local maximum at $\Lambda_m \approx 1.25 \zeta/ \ell$ 
(see Fig.\ \ref{fig1}). Thus, the stationary states exist only in 
a finite interval, $0\leq P \leq P(\Lambda_m)$ and for each value of 
norm in this interval there are two stationary states. This is an 
intrinsic property of the nonlinear Schr{\"o}dinger superlattice 
system. 

Discussing the stability of the stationary states satisfying 
Eq.\ (\ref{eq11}), there are two sources of  instability to be 
considered: longitudinal and transversal perturbations. The 
perturbations of the first type are of the same symmetry with respect 
to transversal degrees of freedom as the stationary states of 
Eq.\ (\ref{eq11}), while the second type of perturbations breaks this 
symmetry. It was shown in \cite{gaid} that stationary states which 
correspond to the branch with $dP / d \Lambda <0$ are unstable due to
the longitudinal perturbations. The states with 
$\Lambda > (4 \zeta^2/3 \ell^2) \sin^2 (\pi/N)$ are, in their
turn, unstable due to the transversal perturbations. 
Thus, one can expect stable stationary solutions for nonlinear 
frequencies satisfying the condition
\begin{equation}
\Lambda \,<\,\frac{\zeta^2}{\ell^2}\,\mbox{min}\{ \frac{4}{3} 
\sin^2(\frac{\pi}{N}),\Lambda_m\} \; . 
\label{eq12}
\end{equation}
In particular, this means that the stationary state 
$\psi_n(z,t)=e^{i\Lambda t}\phi(z)$ can neither exist in the case 
of only one nonlinear layer ($\ell \rightarrow \infty $) nor in 
the quasi-continuum limit ($N \rightarrow \infty$). But in the 
latter case the system supports stationary states which are 
localized in both spatial directions (see Ref.\ \cite{gaid} for 
details).

\section{Discrete NLS models with Long-Range dispersive 
interactions} 

In the main part of the previous studies of the discrete NLS models 
the dispersive interaction was assumed to be short-ranged and a 
nearest-neighbor approximation was used. However, there exist physical 
situations that definitely can not be described in the framework of 
this approximation. The DNA molecule contains charged groups, with 
long-range Coulomb interaction ($1/r$) between them. The excitation 
transfer in molecular crystals \cite{davydov71} and the vibron energy 
transport in biopolymers \cite{scott92} are due to transition 
dipole-dipole interaction with $1/r^3$ dependence on the distance, $r$. 
The nonlocal (long-range) dispersive interaction in these systems 
provides the existence of additional length-scale: the radius of the 
dispersive interaction. We will show that it leads to the bifurcative 
properties of the system due to both the competition between 
nonlinearity and dispersion, and the interplay of long-range 
interactions and lattice discreteness. 

In some approximation the equation of motion is the nonlocal 
discrete NLS equation of the form 
\begin{eqnarray}
i\frac{d}{d t}\psi_n+\sum_{m\neq n}J_{n-m}(\psi_m-\psi_n)+ 
|\psi_n|^2\psi_n=0 \; ,
\label{eq21}
\end{eqnarray} 
where the long-range dispersive coupling is taken to be either 
exponentially, $J_n=J\,e^{-\beta|n|}$, or algebraically, 
$J_n=J\,|n|^{-s}$, decreasing with the distance $n$ between 
lattice sites. In both cases the constant $J$ is normalized 
such that $\sum\limits_{n=1}^{\infty} J_n=1$ for all $\beta$ or 
$s$. The parameters $\beta$ and $s$ are introduced to cover 
different physical situations from the nearest-neighbor 
approximation ($\beta \rightarrow \infty, \; s \rightarrow 
\infty$) to the quadrupole-quadrupole ($s=5$) and dipole-dipole 
($s=3$) interactions. The Hamiltonian 
$H=\sum\limits_{n,m} J_{n-m} |\psi_n-\psi_m|^2-\frac{1}{2} 
\sum\limits_n |\psi_n|^4$, which corresponds to the set of 
equations (\ref{eq21}), and the number of excitations 
$N=\sum\limits_n |\psi_n|^2$ are conserved quantities. 

\begin{figure}[ht]
\begin{center}
\leavevmode
\psfig{figure=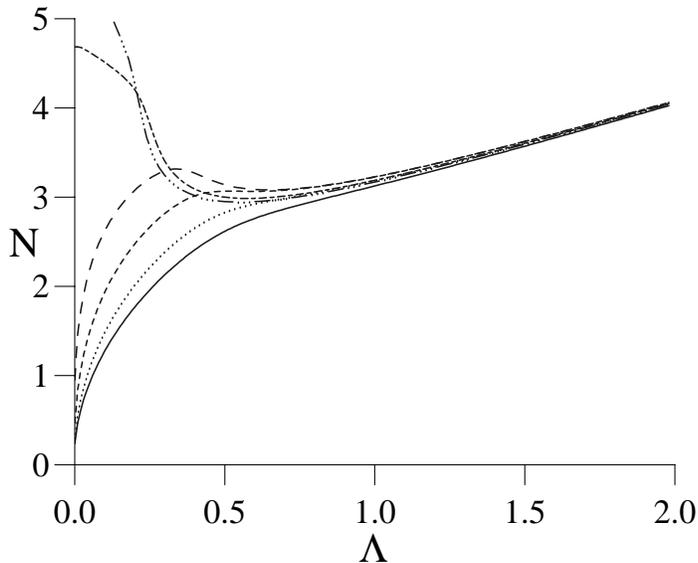,height=3.0in,angle=270}
\caption{Number of excitations, $N$, versus frequency, 
$\Lambda$, found numerically from Eq.\ (\ref{23}) for 
$s=\infty$ (full), 4 (dotted), 3 (short-dashed), 
2.5 (long-dashed), 2 (short-long-dashed),
1.9 (dashed-dotted).}
\label{fig2}
\end{center}
\end{figure} 

We are interested in stationary solutions of Eq.\ (\ref{eq21})
of the form $\psi_n(t)=\phi_n \exp (i \Lambda t)$ with a real shape 
function $\phi_n$ and a frequency $\Lambda$. This gives the 
governing equation for $\phi_n$ 
\begin{equation}
\label{23}
\Lambda \phi_n=  \sum_{m \neq n} J_{n-m} (\phi_m-\phi_n)+ 
\phi_n^{3} \; ,
\end{equation}
which is the Euler-Lagrange equation for the problem 
of minimizing $H$ under the constraint $N=constant$. 

\begin{figure}[h]
\begin{center}
\leavevmode
\psfig{figure=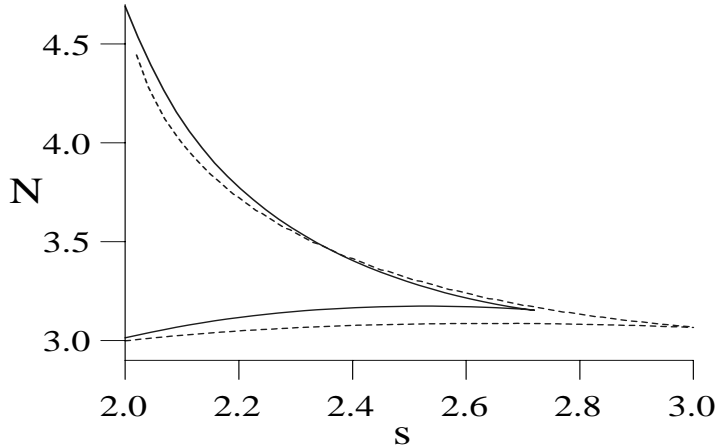,height=2.7in,width=5in,angle=270}
\caption{Shows endpoints of the 
bistability interval for 
$N$ versus dispersion parameter $s$. For $s=s_{cr}$ the endpoints 
coalesce. Analytical dependence (full) gives $s_{cr} \simeq 2.72$. 
Numerical dependence (dashed) gives $s_{cr} \simeq 3.03$.}
\label{fig3}
\end{center}
\end{figure} 

Figure \ref{fig2} shows the dependence $N(\Lambda)$ obtained from 
direct numerical solution of Eq.\ (\ref{23}) for algebraically 
decaying $J_{n-m}$. A monotonic function is obtained only for 
$s>s_{cr}$. For $2<s<s_{cr}$ the dependence becomes nonmonotonic (of 
${\cal N}$-type) with a local maximum and a local minimum. These 
extrema coalesce at $s=s_{cr} \simeq 3.03$. For $s<2$ the local 
maximum disappears. The dependence $N(\Lambda)$ obtained analytically 
using the variational approach is in a good qualitative agreement with 
the dependence obtained numerically (see \cite{gmcr}). Thus the main 
features of all discrete NLS models with dispersive interaction 
$J_{n-m}$ decreasing faster than $|n-m|^{-s_{cr}}$ coincide 
qualitatively with the features obtained in the nearest-neighbor 
approximation where only one stationary state exists for any number 
of excitations, $N$. However in the case of long-range nonlocal NLS 
equation (\ref{eq21}), i.e.\ for $2<s<s_{cr}$, there exist for each 
$N$ in the interval $[N_{l}(s), N_{u}(s)]$ three stationary states 
with frequencies $\Lambda_{1}(N) < \Lambda_{2}(N) < \Lambda_{3}(N)$. 
In particular, this means that in the case of dipole-dipole 
interaction ($s=3$) multiple solutions exist. It is noteworthy that 
similar results are also obtained for the dispersive interaction of 
the exponentially decaying form. In this case the bistability takes 
place for $\beta\,\leq\,1.67$.  According to the theorem which was 
proven in \cite{lst94}, the necessary and sufficient stability 
criterion for the stationary states is $dN /d\Lambda > 0$. Therefore, 
we can conclude that in the interval $[N_{l}(s), N_{u}(s)]$ there 
are only two linearly stable stationary states: $\Lambda_{1}(N)$ 
and $\Lambda_{3}(N)$. The intermediate state is unstable since 
$dN /d\Lambda < 0$ at $\Lambda=\Lambda_2$. 

At the end points ($\Lambda(N_l)$ and $\Lambda(N_u)$) the stability 
condition is violated, since $(dN /d\Lambda)_s$ vanishes. 
Constructing the locus of the end points we obtain the curve that 
is presented in Fig. \ref{fig3}. This curve bounds the region of 
bistability. It is analogous to the critical curve in the van der 
Waals' theory of liquid-vapor phase transition \cite{ll59}. Thus 
in the present case we have a similar phase transition like behavior 
where two phases are the continuum states and the intrinsically 
localized states, respectively. The analog of the temperature is the 
dispersive parameter $s(\beta)$. 

\begin{figure}[h]
\begin{center}
\leavevmode
\psfig{figure=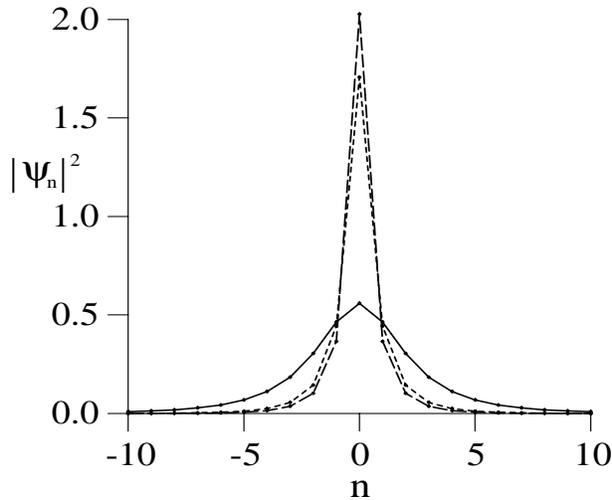,height=3.0in,width=5.0in,angle=270}
\caption{Shapes of three stationary states for $s=2.5$ and 
$N=3.1$. The stable: $\Lambda=0.21$ (full), $\Lambda=0.74$ 
(long-dashed). The unstable: $\Lambda=0.57$ (short-dashed).}
\label{fig4}
\end{center}
\end{figure} 

The shapes of three stationary states in the interval of bistability 
differ significantly (see Fig.\ \ref{fig4}). The low frequency states 
are wide and continuum-like while the high frequency solutions 
represents intrinsically localized states with a width of a few 
lattice spacings. It can be obtained \cite{gmcr} that the inverse 
widths of these two stable states are 
$\alpha_1\, \approx \, \left( N/8J \right)^{1/(s-2)}= 
\left( N/8J \right)^{\ln \xi/(1-2\ln \xi)}, \; \; \; 
\alpha_3\, \approx \, \ln \left( N/J \right)$ 
with $\xi=\exp(1/s)$ being the characteristic length scale of the 
dispersive interaction which is defined as a distance (expressed in 
lattice spacings) at which the 
interaction decreases in two times. It is seen from these expressions 
that the existence of two so different soliton states for one value 
of the excitation number, $N$, is due to the presence of two different 
length scales in the system: the usual scale of the NLS model which 
is related to the competition between nonlinearity and dispersion 
(expressed in terms of the ratio $N/J$ ) and the range of the 
dispersive interaction $\xi$. 

\begin{figure}[h]
\begin{center}
\leavevmode
\psfig{figure=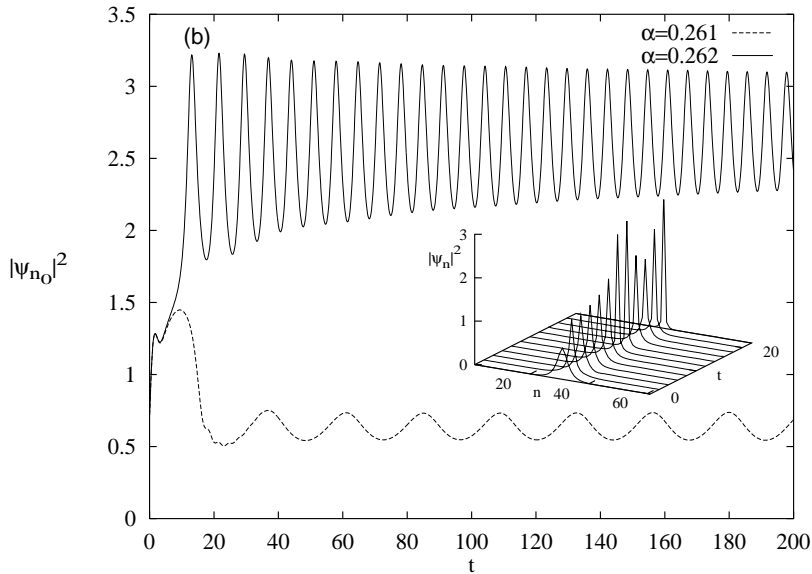,height=3.0in,angle=270}
\caption{Switching from continuum-like to discrete state for 
$\beta=1$. The initial state $\phi_n$ has the frequency 
$\Lambda\simeq 0.31$ and $N=3.6$. The time evolution of 
$|\psi_{n_0}(t)|^2$ when a phase torsion is applied to the center 
site with $\theta=0.261$ (lower curve) and $\theta=0.262$ (upper 
curve), respectively; inset shows time evolution of 
$|\psi_{n}(t)|^2$ for $\theta=0.262$.} 
\label{fig5}
\end{center}
\end{figure} 
\begin{figure}[h]
\begin{center}
\leavevmode
\psfig{figure=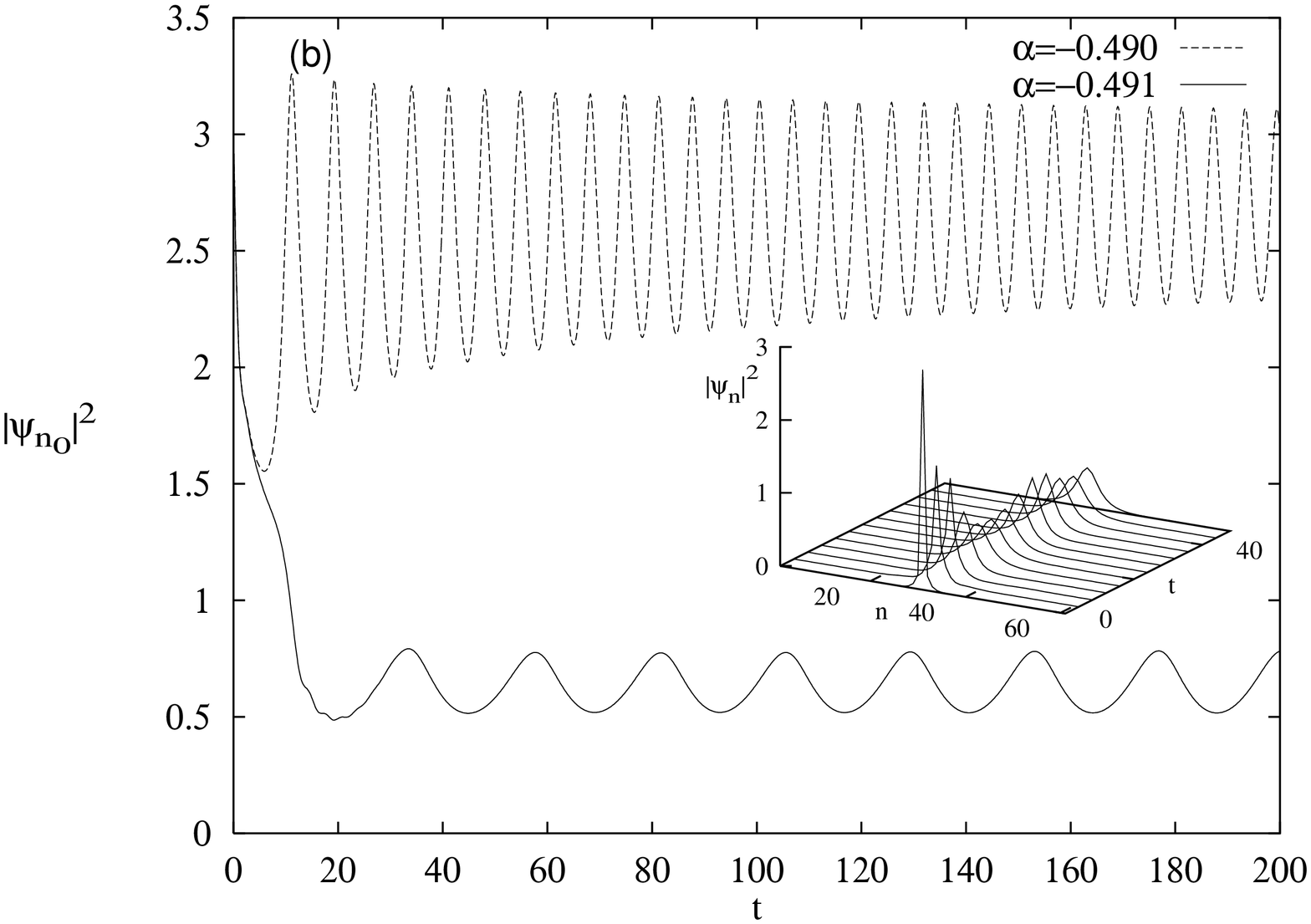,height=3.0in,angle=270}
\caption{Switching from discrete to continuum-like  state for 
$\beta=1$. The initial state $\phi_n$ has the frequency 
$\Lambda\simeq 1.423$ and $N=3.6$. Same as Fig.\ref{fig5} with 
$\theta=-0.490$ (upper curve) and $\theta=-0.491$ (lower curve), 
respectively; inset shows time evolution of $|\psi_{n}(t)|^2$ 
for $\theta=-0.491$ (only a part of a larger system is shown).} 
\label{fig6}
\end{center}
\end{figure} 

Having established the existence of bistable stationary states in 
the nonlocal discrete NLS system, a natural question that arises 
concerns the role of these states in the full dynamics of the model. 
In particular, it is of interest to investigate the possibility of 
switching between the stable states under the influence of external 
perturbations, and to clear up what type of perturbations can 
be used to control the switching. Switching of this type is important 
for example in the description of nonlinear transport and storage of 
energy in biomolecules like the DNA, since a mobile continuum-like 
excitation can provide action at distance while the switching to a 
discrete, pinned state can facilitate the structural changes of the 
DNA \cite{geor96}. As it was shown recently in \cite{mj98}, switching 
will occur if the system is perturbed in a way so that an internal, 
spatially localized and symmetrical mode ('breathing mode') of the 
stationary state is excited above a threshold value. 

We will in sequel mainly discuss the case when the matrix element of 
excitation transfer, $J_{n-m}$, decreases exponentially  with the 
distance $|n-m|$. For $\beta=1$ the multistability occurs in 
the interval $3.23\,\leq\,N\,\leq \,3.78$. It is worth noticing, 
however, that the scenario of switching described below remains 
qualitatively unchanged for all values of $\beta\,\leq\,1.67$, and 
also for the algebraically decaying dispersive coupling with 
$2\,\leq\,s\,\leq\,3.03$.

An illustration of how the presence of an internal breathing mode 
can affect the dynamics of a slightly perturbed stable stationary 
state is given in Figs.\ \ref{fig5} and \ref{fig6}. To excite the 
breathing mode, we apply a spatially symmetrical, localized 
perturbation, which we choose to conserve the number of excitations 
in order not to change the effective nonlinearity of the system. 
The simplest choice, which we have used in the simulations shown 
here, is to kick the central site $n_0$ of the system at $t=0$ by 
adding a parametric force term of the form 
$\theta\delta_{n,n_0}\delta(t)\psi_n(t)$ to the left-hand-side of 
Eq.\ (\ref{eq21}). As can easily be shown, this perturbation affects 
only the site $n_0$ at $t=0$, and results in a 'twist' of the 
stationary state at this site with an angle $\theta$, i.e. 
$\psi_{n_0}(0)=\phi_{n_0}\,e^{i\theta}$. The immediate consequence 
of this kick is, as can been deduced from the form of Eq.\ (\ref{eq21}), 
that $\frac{d}{dt}\left(|\psi_{n_0}|^2\right)$ will be positive 
(negative) when $\theta\,>\,0$ ($\theta\,<\,0$). Thus, we choose 
$\theta\,>\,0$ to obtain switching from the continuum-like state to 
the discrete state, while we choose $\theta\,<\,0$ investigating 
switching in the opposite direction. We find that in 
a large part of the multistability regime there is a well-defined 
threshold value $\theta_{th}$: when the initial phase 
torsion is smaller than $\theta_{th}$, periodic, slowly decaying 
'breather' oscillations around the initial state will occur, while 
for strong enough kicks (phase torsions larger than $\theta_{th}$) 
the state switches into the other stable stationary state. 

It is worth remarking that the particular choice of perturbation is 
not important for the qualitative features of the switching, as 
long as  there is a substantial overlap between the perturbation 
and the internal breathing mode. We believe also that the mechanism 
for switching described here can be applied for any multistable 
system where the instability is connected with a breathing mode.

\section{Stabilization of nonlinear excitations by disorder} 

In this section we discuss disorder effects in  NLS models. Usually 
the investigations of disorder effects have been carried out on 
systems that are integrable - soliton bearing - in the absence of 
disorder. A common argument is that the equations, despite their 
exact integrability, provide a sufficient description of the physical 
systems to display the essential behavior. However, the more common 
physical situation is that integrability, and thus the exact soliton, 
is absent. A relevant example of such an equation is the 
two-dimensional (or higher-dimensional) NLS equation. The two-dimensional 
NLS equation is nonintegrable and possesses an unstable ground state 
solution which, in the presence of perturbations, either collapses or 
disperses (see e.g.\ \cite{af50-a,af50-b}). 

We consider a quadratic two-dimensional lattice with the lattice 
spacing equal to unity. The model is given by the Lagrangian 
\begin{eqnarray}
\label{lagr}
L=\frac{i}{2} \sum_{n,m} \left(\psi_{n,m}^{*}\, \frac{d}{dt} 
\psi_{n,m}-c.c. \right)-H \; , 
\end{eqnarray}
where 
\begin{eqnarray}
H &=& \sum_{n,m}\left(|\psi_{n+1,m}-\psi_{n,m}|^2+|\psi_{n,m+1}- 
\psi_{n,m}|^2 \right. \nonumber \\ 
&-& \left. \frac{1}{2}|\psi_{n,m}|^4 -\epsilon_{n,m}|\psi_{n,m}|^2 
\right)
\label{hamil}
\end{eqnarray}
is the Hamiltonian of the system. In Eqs.\ (\ref{lagr}) and 
(\ref{hamil}) $(n,m)$ is the lattice vector ($n$ and $m$ are integer). 
The first two terms in Eq.\ (\ref{hamil}) correspond to the 
dispersive energy of the excitation, the third term describes a 
self-interaction of the excitation and the fourth term represents 
diagonal disorder in the lattice. Here the random functions 
$\epsilon_{n,m}$ are assumed to have Gaussian distribution with the 
probability $p(\epsilon_{n,m})=\frac{1}{\eta \sqrt{\pi}} 
\exp[-(\epsilon_{n,m}/\eta)^2]$ and have the autocorrelation 
function $\langle \epsilon_{n,m}\epsilon_{n',m'}\rangle =
\eta^2\delta_{n\,n'}\delta_{m\,m'}$, where the brackets 
$\langle .... \rangle$ denote averaging over all realizations of 
the disorder. From the Lagrangian (\ref{lagr}) we obtain the 
equation of motion  for the excitation function in the form 
\begin{eqnarray}
i \frac{d}{dt} \psi_{m,n} &+& (\psi_{m,n-1}+\psi_{m,n+1}+ 
\psi_{m+1,n} +\psi_{m-1,n}-4\psi_{m,n}) \nonumber \\ 
&+& |\psi_{m,n}|^{2}\psi_{m,n}+ 
\epsilon_{m,n}\psi_{m,n}=0 \; . 
\label{deq}
\end{eqnarray}
Equation (\ref{deq}) conserves the norm $N=\sum\limits_{n,m} 
|\psi_{n,m}|^2$ and the Hamiltonian $H$. 

We are interested in the stationary solutions of Eq.\ (\ref{deq}) of 
the form 
\begin{eqnarray}
\psi_{n,m}(t)=\phi_{n,m}\exp(i\Lambda t) \; , 
\label{stat}
\end{eqnarray}
with a real shape function $\phi_{n,m}$ and a nonlinear frequency 
$\Lambda$.

\begin{figure}[h]
\begin{center}
\leavevmode
\psfig{figure=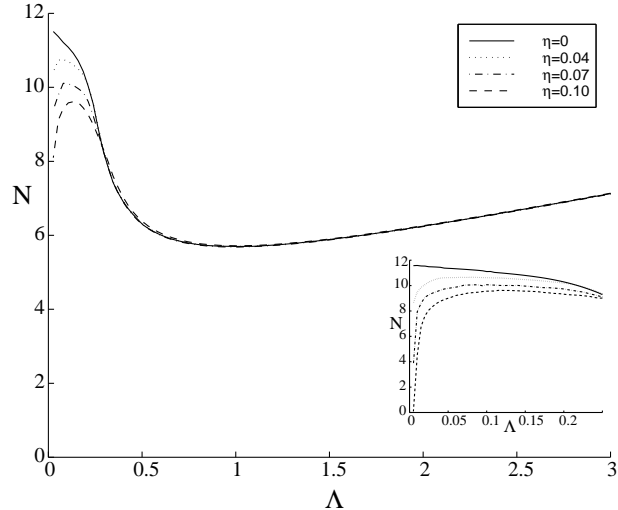,height=3.0in,angle=90}
\caption{The norm $N$ versus nonlinear frequency $\Lambda$ for 
various disorder strengths $\eta$. Homogeneous case $\eta=0$ (solid 
line), $\eta=0.04$ (dotted line), $\eta=0.07$ (dashed-dotted line) 
and $\eta=0.1$ (dashed line).} 
\label{fig7}
\end{center}
\end{figure} 

Equation (\ref{deq}) together with Eq.\ (\ref{stat}) constitute a 
nonlinear eigenvalue problem which can be solved numerically using 
the techniques described in Ref. \cite{af89}. The dependences 
$N(\Lambda)$ in the absence and in the presence of disorder are 
shown in Fig.\ \ref{fig7}. It has previously been shown 
\cite{af158,af159,lst94,af47} that the linear stability of the 
stationary states in the discrete case is determined by the condition 
$dN/d\Lambda>0$. Thus, in the case without disorder (solid curve in 
Fig.\ \ref{fig7}) the low-frequency ($0\leq \Lambda\leq 
\Lambda_{min}=1.088$) nonlinear excitations in the discrete 
two-dimensional NLS model are unstable. It is important that in 
the continuum limit ($\Lambda\rightarrow 0$) the norm $N(\Lambda)$ 
tends to the non-zero value $N_c\simeq 11.7$.

Other lines in Fig.\ \ref{fig7} show the dependence $N$ on 
$\Lambda$ for the stationary solutions of Eq.\ (\ref{deq}) in the 
presence of disorder. The results have been obtained as averages 
of $150$ realizations of the disorder. Several new features arise 
as a consequence of the disorder. In the continuum limit 
($\Lambda\rightarrow 0$) we no longer have $N=N_c$ with 
$dN/d\Lambda = 0$. Instead we have $N\rightarrow 0$ with 
$dN/d\Lambda >0$ signifying that the disorder stabilizes the 
excitations in the low-frequency limit. The disorder creates a stability 
window such that a bistability phenomenon emerges. Consequently there 
is an interval of the excitation norm in which two stable excitations 
with significantly different widths have the same norm. 

Furthermore, we see that the disorder creates a gap at small 
$\Lambda$ in which no localized excitations can exist, and that the 
size of this gap apparently is increased as the variance of the 
disorder is increased. It is also clearly seen that as $\Lambda$ 
increases (decreasing width) the effect of the disorder vanishes 
so that the very narrow excitations are in average unaffected by the 
disorder. It is important to stress that this is an average effect, 
because for each realization of the disorder the narrow excitation will 
be affected. The narrow excitation will experience a shift in the 
nonlinear frequency equal to the amplitude of the disorder at the 
position of the excitation. 

\begin{figure}[h]
\begin{center}
\leavevmode
\psfig{figure=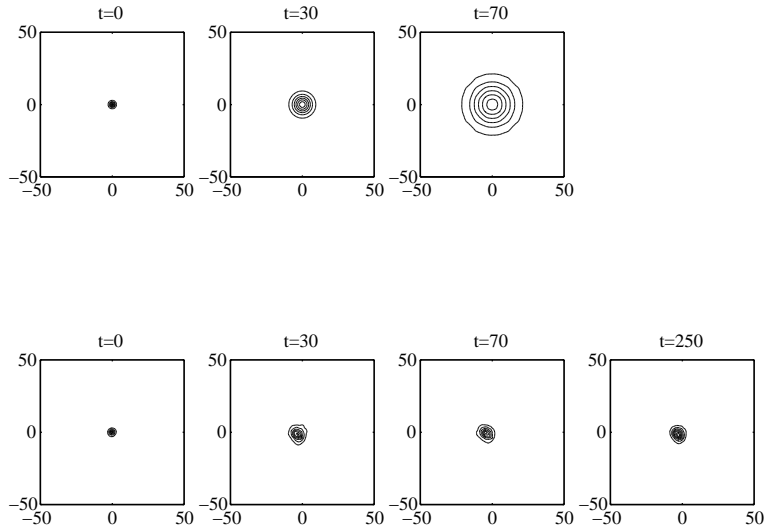,height=3.0in,angle=90}
\caption{Evolution of an initial excitation of the norm $N=10.4402$ 
without disorder (upper part) and with disorder strength $\eta=0.1$ 
(lower part).}
\label{fig8}
\end{center}
\end{figure} 

The bistability we observe in Fig.\ \ref{fig7} occurs due to the 
competition between two different length scales of the system: one 
length scale being defined by the relation between the nonlinearity 
and the dispersion, while the other length scale being defined by 
the disorder. A similar effect was observed by 
Christiansen {\em et al.} \cite{cg97} for the one-dimensional 
discrete NLS equation with a quintic nonlinearity. The latter is 
quite natural because as it is well known (see e.g. \cite{ob94}) 
the properties of the two-dimensional NLS model with a cubic 
nonlinearity are similar to the properties the one-dimensional NLS 
equation with a quintic nonlinearity. 
 
Having studied the stationary problem it is vital to compare the 
results to full dynamical simulations. Therefore we carry out a 
numerical experiment launching a pulse in a system governed by 
Eq.\ (\ref{deq}). Specifically, stationary solutions (\ref{stat}) of 
Eq.\ (\ref{deq}) with $\Lambda=0.14$ (after reducing the amplitude 
of these solutions by $5$\%) were used as initial conditions of 
the dynamical simulations. Examples of the described experiment are 
shown in Fig.\ \ref{fig8}. As is seen the pulse behavior in the 
absence of disorder and in the presence of disorder (we presented here 
a realization corresponding to the disorder variance $\eta=0.1$) 
differs drastically. While the pulse rapidly disperses in the ideal 
system (the contour plot for $t=250$ is absent because the pulse 
width is of the system size), the process is arrested in the 
disordered system. After some transient behavior the excitation 
stabilizes and attains an approximately stationary width. The dynamical 
simulations thus support the conclusion that otherwise unstable 
excitations are stabilized by the presence of disorder in the low 
frequency limit. 

Analytical theory of soliton states in disordered NLS models based 
on the collective coordinate approach and on  the Rice's theorem 
from the theory of random processes \cite{kree} is presented in 
Refs.\ \cite{cg97} and \cite{gh98}.

\section{Summary} 

In summary we have shown that the presence of competing length scales 
leads to  multistability phenomena in nonlinear Schr{\"o}dinger 
models. We have analyzed three types of the NLS models. The nonlinear 
Schr{\"o}dinger-Kronig-Penney model presents an example where two 
competing length scales exist: the width $\zeta$ of the soliton in the 
nonlinear Schro{\"o}dinger equation and the interlayer spacing $\ell$. 
Due to the interplay between these two length 
scales the localized stationary states exist only in a finite 
interval of the excitation power. Two branches of stationary states 
exist  but only the low-frequency branch is stable. 

In discrete nonlinear Schr{\"o}dinger models with long-range 
dispersive interactions there exist three types of length scales: 
the soliton width, the lattice spacing and the radius of the 
dispersive interaction. Here the competition of the length scales 
provides the existence of three branches of stationary states. 
Two of them: low-frequency branch which contains continuum-like 
excitations and high-frequency one with intrinsically localized 
excitations, are stable. It is shown that a controlled switching 
between narrow, pinned states and broad, mobile states is possible. 
The particular choice of perturbation is not important for the 
qualitative features of the switching, as long as  there is a 
substantial overlap between the perturbation and the internal 
breathing mode. The switching  phenomenon could be important for 
controlling  energy storage and transport in DNA molecules. 

Considering nonlinear excitations in two-dimensional discrete 
nonlinear Schr{\"o}dinger models with disorder it was found that 
otherwise unstable continuum-like excitations can be stabilized by 
the presence of the disorder. For the very narrow excitations 
the disorder has no effect on the averaged 
behavior. Bistability which was observed in this case is very similar 
to the bistability that occurs in nonlocal NLS models. Here the 
bistability arises on similar grounds because of competition between 
the solitonic length scale and the length scale defined by the 
disorder.

\acknowledgments 

Yu.B.G.\ thanks MIDIT, the Technical University of Denmark for the 
hospitality. Yu.B.G.\ and S.F.M.\ acknowledge support from the 
Ukrainian Fundamental Research Fund under grant 2.4/355. 

\bibliographystyle{prsty} 
\bibliography{scale} 

\end{document}